\documentclass[12pt]{amsart}
\usepackage{graphics}
\usepackage{amsthm}
\usepackage{amssymb}
\usepackage{amsmath}
\usepackage{amsfonts}
\usepackage{epic}
\usepackage{curves}
\usepackage{epsfig}
\usepackage{bm}
\usepackage{amscd}
\input{psfig.sty}

\newcommand{\baseRing}[1]{\ensuremath{\mathbb{#1}}}
\newcommand{\Z}{\baseRing{Z}}
\newcommand{\C}{\baseRing{C}}
\newcommand{\N}{\baseRing{N}}
\newcommand{\R}{\baseRing{R}}

\theoremstyle{plain}

\theoremstyle{definition}

\def\be{\begin{equation}}
\def\ee{\end{equation}} 
\title{Remarks on free field realization of $SL(2,\R )_k /U(1) \times U(1)$ WZNW model}

\author{Gast\'on E. Giribet}
\address{G.G. Institute for Advanced Study. Einstein Drive, Princeton NJ08540. On leave from University of Buenos Aires.}

\author{Daniel E. L\'opez-Fogliani}
\address{D.L. IFT and Departamento de F\'isica Te\'orica C-XI, Universidad Aut\'onoma de Madrid. Cantoblanco 28049, Madrid, Spain.}
\thanks{Pre-print numbers: FTUAM 04/09, IFT-UAM/CSIC-04017.}

\begin{document}


\begin{abstract}

Free field representations of vertex algebra in $SL(2,\R)_k /U(1) \times U(1)$ WZNW model are constructed by considering a twisted version of the Bershadsky-Kutasov free field description of discrete states in the two-dimensional black hole CFT. These correspond to conjugate representations describing primary states in the model on $SL(2,\R)/U(1) \times U(1)$. A particular evaluation of these leads to identities due to the spectral flow symmetry of $\hat {sl(2)} _k$ algebra.

The computation of correlation functions violating the $\omega $inding number conservation is discussed and, as an application, these are compared with analogous results known for the sine-Liouville theory. Exact agreement is observed between both analytic structures.



\end{abstract}


\maketitle


\section{Introduction}
 
Among many interesting applications of this model, the $SL(2,\R )_k/U(1)\times U(1)$ WZNW can be identified as the CFT describing the string theory in $AdS_3$.

In reference \cite{nos2}, the free field realization of string theory in $AdS_3$ was studied in terms of the WZNW model formulated on this manifold, and it was shown there how the $\omega $inding sectors of Hilbert space naturally appear in such construction. The Coulomb gas integral representation was analyzed in this context and used to explicitly compute correlation functions \cite{nos3} describing string scattering processes. The free field representation of $SL(2,\R )_k$ WZNW model was also studied in \cite{ads3}; see references therein.

Here, with the intention to advance in the understanding of the free field representation of this non-rational CFT, we present an extension of the results of references \cite{nos2,nos3,fh}. Precisely speaking, the points treated here continue our previous study \cite{nos2,nos3} about the Dotsenko integral representation of correlation functions in Wess-Zumino-Novikov-Witten (WZNW) model formulated on the product between the quotient $SL(2,\R )/U(1)$ (which describes string theory on the euclidean version of two-dimensional black hole \cite{witten}) and a time-like free $U(1)$ boson.



To begin, let us motivate the topic by mentioning that the interest of this particular conformal model is mainly based in the following two aspects: First, the successful description of correlation functions in $AdS_3$ string theory in terms of the free field representation, which nourishes the intention to study the feasibility and fruitfulness of such formalism in this and other conformal models in more detail. Secondly, the conformal theory on the product $SL(2,\R )/U(1) \times U(1)$, by itself, turns out to be closely related with many other interesting systems in string theory: $e.g.$ it is related with the N=2 Kazama-Suzuki coset models, with the $\hat c=1$ conformal theory and, of course, with black hole geometries in two and three dimensions. See for instance the interesting works on related subjects \cite{eguchi,samir}.

As mentioned, the interactions of $\omega $inding\footnote{We will use the name of $\omega $inding strings to refer to the spectral flow degree of freedom, even though its geometrical meaning is not strictly a winding number for the short string states.} strings in $AdS_3$ (accordingly with the construction proposed in \cite{mo}) were described \cite{nos3} in terms of the WZNW model formulated on the mentioned product manifold. More precisely, it was shown how the correlation functions can be computed in terms of the free field description of this theory beyond the near boundary limit by means of the analytic continuation of certain integral equations (see also \cite{ondas}). Three-point functions representing string scattering processes violating the $\omega $inding number conservation were explicitly calculated for the cases where two of the three interacting states were represented by highest-weight vectors of the $SL(2,\R )$. In fact, one of the preliminary goals of this note is to extend the free field computation presented in \cite{nos3} in order to observe how {\it the group theoretical factor}, which parametrizes the discrete representations of $SL(2,\R )$, appears in a more general case. This will enable us to study the relation existing with analogous correlators presented in the literature for the case of sine-Liouville conformal field theory \cite{fh}. 


This note is organized as follows: First, in section 2, we briefly review the results of reference \cite{nos2, nos3} and, consequently, we extend in section 3 the conjugate representations of $\hat {sl(2)} _k$ vertex algebra in order to describe, for example, the $m$-dependent group theoretical factor standing in the computation of three-point function for $\omega $inding violating processes. We remark important aspects of the Dotsenko conjugate representations which were not addressed in our previous works in the subject. Besides, in order to avoid redundances, we refer to the mentioned works for the details.

The free field realization leading to the integral representation of two and three-point correlators in WZNW model formulated on $SL(2,\R)/ U(1) \times U(1)$ is reviewed and, consequently, previous discussions about the explicit examples of realizations of the $\hat {sl(2)} _k$ operator algebra are substantially extended. For instance, one of our main goals is to show that the free field representation of the simplest discrete states in the two-dimensional black hole admits an extension which leads to construct conjugate representations of Kac-Moody primary states on the product manifold $SL(2,\R )/U(1)\times U(1)$.

Then, the computations of two and three-point functions are revisited by using this conjugate representations for vertex operators. Explicit formula for $\omega $inding violating scattering processes is written down. As an application, we study in section 5 the construction of correlators representing scattering processes violating the $\omega $inding number conservation in sine-Liouville theory, which has been conjectured to be a dual model of the string theory on the {\it cigar} manifold $SL(2,\R )/U(1)$. This will enable us to work out an extension of the comparison
effectuated in \cite{fh} regarding the pole structure of both sine-Liouville
model and the non-compact WZNW theory. We show here that the case of violating
$\omega $inding number (let us denote it $\delta \omega =1$) in both conformal models presents similar degree of agreement as the comparison previously effectuated in \cite{fh} for the particular (conservative) case $\delta \omega =0$. Then, the agreement between the analytic structure of both CFT's at the level of three-point correlators seems to be exact. The group theoretical factor arising in the formula for WZNW model and the integration over the zero-modes  of sine-Liouville are crucial points for asserting such agreement.



\section{Free field representation}

This section and the following are devoted to a discussion on the free field representation of $\hat {sl(2)} _k$ vertex operators. We extend the results of references \cite{nos2} and \cite{nos3}; we refer to those papers for the details. 

\subsection{The action and conformal field theory}

Let us start by reviewing the free field representation of the CFT. In \cite{nos3}, the construction of the conformal model on $SL(2,\R )/U(1) \times U(1)$ was achieved by means of the realization of the coset manifold $SL(2,\R )/U(1)$ presented in the quoted references \cite{dvv} and \cite{bk}. The strategy of those works was to realize the theory formulated on the $SL(2,\R )/U(1)$ black hole by introducing an additional space-like bosonic field $X(z)$ describing the $U(1)$ subgroup which, combined with a $(b,c)$ ghost system of spin $(1,0)$, allows to gauge out the Cartan element. Our construction \cite{nos2} includes a new time-like scalar field $T(z)$ in order to describe the product space $SL(2,\R )/U(1) \times U(1)$ which naturally leads to parametrize the $\omega $inding sectors of the Hilbert space of string theory on $AdS_3$. This auxiliary $U(1)$ factor turns out to be useful to realize the theory on the coset in terms of a convenient free field representation.

Then, the (quantum corrected) action of the conformal model combines the non-linear $\sigma$-model yielding from string theory on the $AdS$ background configuration with the inclussion of these auxiliary free fields; namely
\begin{equation}
S_1 =\frac 1{4\pi }\int d^2z\left(
\frac 12 \partial \phi \bar \partial \phi -\sqrt {\frac {2}{k-2}}R\phi +\beta \bar
\partial \gamma +\bar \beta \partial \bar \gamma + \partial X \bar \partial X - \partial T \bar \partial T \right) \nonumber +S_{I} \label{laccion}
\end{equation}
where\footnote{the contribution due to the $(b,c)$ ghost system is not explicitly written here.}
\begin{eqnarray}
S_{I}=\frac {M }{4\pi }\int d^2z {\mathcal L} _I (z,\bar z) \ , \ \ \ {\mathcal L} _I =  \beta \bar \beta e^{-\sqrt{\frac {2}{k-2}}\phi }  \label{arriba}
\end{eqnarray}
In the Coulomb gas realization of the correlators, the interaction term (\ref{arriba}) acts as the insertion of screening operators. The parameter $M$ represents the black hole mass in two dimensions \cite{dvv,witten}, even tough its relevance for physical meaning is only encoded in its sign.

It was demostrated in \cite{nos3} that two and three-point correlation functions defined by this action exactly agree with the analogous computation effectuated by replacing the interaction term $S_I$ by
\begin{eqnarray}
\tilde{S}_{I} &=&\frac {\tilde M }{4\pi }\int d^2z \left( {\mathcal L} _I (z,\bar z) \right) ^{k-2}  \label{arribasss}
\end{eqnarray}
if certain precise relation between the parameters $M$ and $\tilde {M}$ holds (see also \cite{andreev}).

This equality reflects a class of duality between strong and weak regimes ($k-2 \leftrightarrow 1/k-2$), even though it is not an explicit symmetry of the lagrangian formulation of WZNW model, {\it i.e.} since the background charge (dilaton term) is not manifestly invariant under the replacement $k-2 \rightarrow 1/k-2$; see below. This symmetry seems to be valid for this CFT and it turns out to be an important aspects because of the existence of a similar duality in quantum Liouville field theory. Unlike the theory on $SL(2,\R )$, in Liouville CFT the strong-weak duality manifestly appears as a symmetry of the quantum action $S_1 - S_I$, $i.e.$ reflected in the interchange between $b$ and $b^{-1}$ (following the usual nomenclature; {\it e.g.} see \cite{teschnerliouville}, where the importance of both interaction terms in the case of the Liouville model was pointed out).

Of course, the basic form of the relation holding between the parameters $M$ and $\tilde {M}$ can be infered from the KPZ scaling behavior of the coupling constants; notice that, by performing the global transformation
\begin{eqnarray*}
\phi \rightarrow \phi +\phi _0 \quad \quad \gamma \rightarrow e^{-\varphi _0 }\gamma \quad \quad \beta \rightarrow e^{\varphi _0 }\beta ,
\end{eqnarray*}
which is a symmetry of the free field action $S_1-S_I$, the following relations are obtained
\begin{eqnarray}
\frac {\delta M}{M} = \frac {1}{k-2}\frac {\delta \tilde{M}}{\tilde {M}}
\end{eqnarray}
By integrating these, we can write
\begin{equation}
\tilde{M}=f_1 (k-2) M ^{k-2}
\end{equation}
for certain function $f_1 (k-2)$, which satisfies $\lim _{\epsilon \rightarrow 1} |f_1 (\epsilon )|=1$. A direct computation \cite {nos3} leads to obtain the exact expression $f_1 (x) = \frac {1}{\pi } \gamma \left( 1-x \right) \left( \pi \gamma \left( x^{-1} \right)  \right) ^{x}$, being $\gamma (x)= \Gamma (x) / \Gamma (1-x)$. This satisfies $f_1 ^{x} (x^{-1}) f_1 (x)=1$.

If the interaction term (\ref{arribasss}) is used (instead (\ref{arriba})) to compute the $N$-point correlators as insertions of $s$ screening charges in Coulomb gas prescription, then the divergences in correlation functions (similar to those which yield from the non-compactness of target space in Liouville theory) would appear due to a $\Gamma(-s)$ factor standing in the integration over the zero-mode $\phi _0$. Some of these divergences, those which are located at $\sum _{i=1}^{N} j_i =k-3$, are analogous to the pole conditions conjectured\footnote{notice that the substitution $j \rightarrow j-1$ has to be considered in order to conciliate the nomenclature with the one used in \cite{mo3}; see footnote 19 of arXiv:hep-th/011180v2.} in \cite{mo3} and interpreted as instantonic contributions to $N$-point functions in $AdS_3$. This agreement suggests an identification between this pole structures and the instantonic contributions refered in \cite{mo3}. Then, this provides us an argument for asserting that, indeed, these poles do occur in $N$-point correlation functions after the integration over the $(x,\bar x)$ $SL(2,\R)$-isospin coordinates.


\subsection{Wakimoto free field representation of $\hat {sl(2)}_k$}

The WZNW theory on $SL(2,\R) /U(1) \times U(1)$ admits a realization in terms of free fields\footnote{The free field description was extensively studied in the literature; we refer for instance to references \cite{enganchados1}.}.

The stress-tensor of the conformal model on the product $SL(2,\R )/U(1) \times U(1)$ is given by
\begin{equation}
{\mathcal T}=\beta \partial \gamma -\frac 12(\partial \phi )^2-\frac 1{\sqrt{2k-4}}\partial ^2\phi -\frac 12(\partial X)^2-b\partial c+\frac 12(\partial
T)^2  \label{cacatua8}
\end{equation}
and the central charge is then given by
\begin{eqnarray*}
c=3+\frac {6}{k-2} ,
\end{eqnarray*}
It corresponds to the Sugawara stress-tensor, contructed from the $\hat {sl(2)} _k$ Kac-Moody currents
\begin{eqnarray}
J^{+}(z) &=&\beta (z)  e^{i\sqrt{\frac{2}{k}}(X(z)+T(z))}  \label{laconc} \\
J^{3}(z) &=&-\beta (z)\gamma (z)-\sqrt {\frac{k-2}{2}}\partial \phi
(z) -i\sqrt{\frac{k}{2}} \left( \partial X(z) +\partial T(z) \right)  \\
J^{-}(z) &=&\left( \beta (z)\gamma ^{2}(z)+\sqrt{2k-4}\gamma
(z)\partial \phi (z)+k\partial \gamma (z) \right)  e^{-i\sqrt{\frac{2}{k}}(X(z)+T(z))} \label{hadetumadre}
\end{eqnarray}
and taking into account the contribution of the spin $(1,0)$ auxiliary ghost system; see \cite{nos3,dvv,bk,becker} for details of construction.
Thus, the theory admits a representation of the current algebra in terms of free fields; namely an scalar field $\phi $ with background charge, a $(\beta ,\gamma )$ ghost system and the two additional free bosons $X$ and $T$. These fields satisfy the following correlators
\[
\left< \phi (z) \partial \phi (w)\right> = \left< X (z) \partial X (w)\right> = \left< \partial T(z) T (w)\right> = \left< \beta (z) \gamma (w)\right> = \frac {1}{z-w}
\]
The vertex operators in this CFT are given by Virasoro primaries $\Phi ^{\omega }_{j,m,\bar m} $ creating $SL(2,\R )_k$ states $\left| \Phi _{j,m,\bar m} ^{\omega } \right>$ by acting on the invariant vacuum $\left| 0\right >$; namely
\begin{eqnarray}
\lim _{z\rightarrow 0}\left( {\Phi}^{\omega }_{j,m,\bar m} (z)+R (j,m) {\Phi}^{\omega }_{-1-j,m,\bar m} (z) \right) \left| 0 \right> \ = \left| \Phi _{j,m,\bar m} ^{\omega} \right> \label{pou8}
\end{eqnarray}
where
\begin{equation}
R (j,m)= B(j)\frac{\Gamma (1+j+m)}{\Gamma (-m-j)}\frac {\Gamma (1+j-m)}{\Gamma (m-j)} \frac{\Gamma (-2j-1)}{\Gamma(2+2j)}  \label{efffe}
\end{equation}
and where $B(j)$ is given by the two-point function (see below (\ref{herre})).

The large $\phi $ limit of these operators can be realized by the Wakimoto free field representation as follows
\begin{eqnarray}
{\Phi}^{\omega }_{j,m,\bar m}=: \gamma ^{j-m}e^{j\sqrt{\frac{2}{k-2}}\phi }e^{i\sqrt{%
\frac 2k}mX}e^{i\sqrt{\frac 2k}(m+\frac k2 \omega )T} :  \times  \ h.c. \label{asdf1}
\end{eqnarray}
where $h.c.$ refers to the anti-holomorphic part, which is similar. These operators satisfy the $\hat {sl(2)} _k$ block structure of Kac-Moody primaries, which is encoded in the following operator product expansion 
\begin{eqnarray}
J^ 3 (z) \Phi ^{0}_{j,m,\bar m} (w) &=& \frac {m}{(z-w)}  \Phi ^{0}_{j,m,\bar m}(w) + ... \\ J^ {\pm} (z) \Phi ^{0}_{j,m,\bar m} (w) &=& \frac {\mp j - m}{(z-w)}  \Phi ^{0}_{j,m \pm 1,\bar m} (w)+ ...
\label{g}
\end{eqnarray}
And, analogously, operators (\ref{pou8}) for generic $\omega $ are actually Kac-Moody primaries with respect to the algebra generated by the currents obtained by acting with the spectral flow automorphism on (\ref{laconc})-(\ref{hadetumadre}). 

The action of spectral flow\footnote{The spectral flow symmetry was studied in different context in the literature; let us mention the interesting works \cite{enganchados2}.} on the modes $J^a_n = \frac {1}{2 \pi i} \oint dz J^a (z)z^n $ is defined as follows
\begin{eqnarray*}
J^{\pm}_n \rightarrow J^{\pm}_{n\pm \omega} \ , \ \ J^{3}_n \rightarrow J^{3}_{n} - \frac k2 \omega \delta _{n,0}
\end{eqnarray*}
Then, it is feasible to see that operators (\ref{pou8}) have conformal dimension
\begin{equation}
h _{(j,m)} =-\frac {j(j+1)}{k-2} +\frac {m^2}{k}-\frac {(m+\frac {k}{2}\omega)^2}{k}  \label{msc}
\end{equation}
where $j$ and $m$ parametrize the (universal covering of) representations of $SL(2,\R)$, as usual. This reproduces the spectrum of string theory on $AdS_3$, where the angular momentum, the energy and the $\omega $inding number of string states are given by the quantities $p=m-\bar m$, $e=m+\bar m +k\omega $ and $\omega $ respectively. It has to be contrasted with the case of the cigar manifold, where the $\omega $inding modes are characterized by the quantity $k\omega =m+\bar m$ while the difference $p=m-\bar m$ represents the angular momentum in the asymptotic cylinder. The relation between the three-dimensional Anti-de Sitter space and the product between the two-dimensional euclidean black hole and a time-like coordinate was studied in \cite{mo3} and \cite{hw}. The main distinction between $AdS_3$ and the cigar geometry consists in the respective dispersion relation between the $\omega $inding modes and the $e$nergy of both models. Actually, the spectrum of the theory on the coset corresponds to the restriction of vanishing energy in $AdS_3$.

Notice that the mass-shell condition (\ref{msc}) remains invariant under the transformations $(j,m,\bar m,\omega ) \rightarrow (j,-m,-\bar m,-\omega )$, $(j,m,\bar m,\omega ) \rightarrow (-1-j,m,\bar m,\omega )$, $(j,m,\bar m,0 ) \rightarrow (j,\mp m,\pm \bar m,0 )$ and $(j,j,j,0) \rightarrow (-\frac k2 - j,\frac k2 + j,\frac k2 + j,-1)$. These and their compositions correspond to symmetries of the spectrum which are translated into identities between different representations of $SL(2,\R )_k$.

It is clear that by excluding the auxilliary fields $X$ and $T$ (or, which is equivalent, by restricting to the sector $\omega =0$) in the free field realization of $J^a$ one finds the standard Wakimoto representation of $\hat {sl(2)} _k$. With the purpose to make the $sl(2,\R )$ structure of representation (\ref{laconc})-(\ref{hadetumadre}) more clear, let us consider the functional form $\Phi _{F_a} = F_a(\gamma ) e^{\sqrt {\frac {2}{k-2}} j  \phi}$ and the differential operators 
\[
{\mathcal D} ^+ = \frac {\partial }{\partial \gamma } \ , \ \ {\mathcal D} ^3 = -\gamma \frac {\partial }{\partial \gamma } + j \ , \ \ {\mathcal D} ^- = \gamma ^2 \frac {\partial }{\partial \gamma } -2j \gamma 
\]
which form a representation of $sl(2)$. Then, by {\it reading} the simple pole of the OPE between the currents $J^a$ and operators $\Phi _{F^a}$ we can translate the problem of findind a representation ${\Phi _{F^a}}$ diagonalizing $J^a$ into the problem of finding eigenfunctions $F^a$ of the operators ${\mathcal D} ^a$. Then, the solutions to this problem are given by 
\[
F^3 _{j,m} (\gamma ) = \gamma ^{j-m} \ , \ \ \ F^{\pm} _{j,m} (\gamma ) = \gamma ^{j \mp j } e^{\pm m \gamma ^{\pm 1}}
\]
where $F^3 _{j,m} (\gamma )$ corresponds to the standard base (\ref{asdf1}).

\section{Conjugate Representations}

\subsection{Conjugate representations for $SL(2,\R )_k /U(1) \times U(1)$ WZNW}

Conjugate representations should be also considered in order to realize the correlation functions \cite{dot3}. We state here that these can be constructed by starting from the free field representation of certain discrete states appearing in the string spectrum on two-dimensional black hole manifold (signaled by Bershadsky and Kutasov in reference \cite{bk}). In order to do this, it is necessary to perform the change $(j,m,\bar m) \rightarrow (-j-\frac k2 , m-\frac k2, \bar m -\frac k2)$ in the states of the coset $SL(2,\R )/U(1)$. This corresponds to a {\it twisting} of the states on the product $SL(2,\R )/U(1)\times U(1)$ since the contribution ({\it i.e.} the charge) of the time-like field $T$ remains invariant in this procedure; namely
\begin{equation}
\tilde {\Phi}^{\omega }_{j,m,\bar m}=\frac{(-1)^{m-j}}{\Gamma (m+j+1)} :\beta ^{j+m}e^{-\sqrt {\frac {2}{k-2}}(j+\frac k2)\phi }e^{i\sqrt{\frac 2k}%
(m-\frac k2)X} e^{i\sqrt{\frac 2k}(m+\frac k2 \omega)T} :  \times  \ h.c.  \label{conj+-}
\end{equation}
The normalization here turns out to be the adequate to realize the standard block structure of the $\hat {sl(2)}_k$ current algebra (\ref{g}) and, consistently, coincides with the one leading to the group theoretical factor in the correlators. Certainly, it is instructive to verify that (\ref{conj+-}) satisfies the $\hat {sl(2)} _k$ block structure of the Kac-Moody primaries of the representation $-1-j$ with respect of the generators obtained by acting with the spectral flow with parameter $\omega $ on the modes of the currents (\ref{laconc})-(\ref{hadetumadre}) ({\it i.e.} $\Phi _{-1-j,m,\bar m}^{\omega } \sim \tilde {\Phi}^{\omega }_{j,m,\bar m}$). Moreover, we could also consider, instead (\ref{conj+-}), a normalization factor $\frac{(-1)^{j+m}B(j)\Gamma (-2j-1)}{\Gamma (m-j) \Gamma (2j+2)}$ yielding to representations\footnote{we are omitting here the complex conjugate contribution, which is similar; for instance, operators (\ref{asdf1}) present a factor $\sim e^{i\sqrt {\frac 2k} (mX_R(z)+\bar {m} X_L(\bar {z}))}$ and analogously for the other free fields, ({\it cf.} (\ref{fuld}) below). A minor differens appear between holomorphic and anti-holomorphic parts since the phases in the normalization in (\ref{conj+-}) are $(-1)^{m-j}$ and $(-1)^{-\bar m-j}$ respectively.} $\tilde {\Phi } ^{\omega }_{j,m,\bar m} \sim \Phi ^{\omega }_{j,m,\bar m}$.

Restricted to highest-weight states, this correspondence between quantum numbers $j$ and $-\frac k2 -j$ is associated to the fact that the spectral flow automorphism in the $\omega =1$ sector is\footnote{which is basically defined by the application $J^{\pm}_n \rightarrow J^{\pm}_{n\pm 1}$ and $J^{3}_0 \rightarrow J^{3}_{0} - \frac k2 $.} closed among the standard representations of $SL(2,\R )$ and simply maps highest (lowest)-weight states of the certain representation $j$ in lowest (resp. higest)-weight states of the representation $-\frac k2 -j$.

An important point that deserves to be remarked is the fact that the functional form (\ref{conj+-}) represents Kac-Moody primary fields indeed, and this generalizes the highest-weight form used in references \cite{nos2,nos3,dot1,petko,petko2}. Notice that (\ref{conj+-}) also includes, as the particular case $\tilde {\Phi }^{-1} _{j+\frac k2-1,j+\frac k2,j+\frac k2}$, the representations introduced originally by Dotsenko in \cite{dot1} in the context of $SU(2)_k$ WZNW model.

Thus, the free field representation of the discrete states in the coset $SL(2,\R )_k/U(1)$ admits an extension to the theory on the product $SL(2,\R )_k /U(1)\times U(1)$, leanding to conjugate representations of Kac-Moody primary states. Indeed, this extension, despite its simplicity, has non trivial implications since, as we know, the discrete states on the coset do not represent Kac-Moody primary fields, but descendents states in the Verma modulo. Actually, the inclusion of the additional scalar fields and the twisting of both $U(1)$ charges turn out to be the reason of the presence of primary operators with such $\beta $-dependent particular functional form for primary states. This is an example of the fact that, even though the theory on $SL(2,\R)/U(1)$ and the theory on $SL(2,\R)/U(1) \times U(1)$ admit similar realizations, these present very distinctive properties in the spectrum.

We summarize the properties of this conjugate representations $\tilde {\Phi }^{\omega }_{j,m,\bar m}$ in the concluding remarks.

\subsection{Spectral flow symmetry and conjugate representations}

On the other hand, the identity between different (discrete) representations of $SL(2,\R)$, due to the spectral flow symmetry of $\hat {sl(2)}_k$, allows us to write down a new conjugated representation of highest-weight states; namely
\begin{equation}
\hat {\Phi } _{j,j,j}^{\omega } = : \gamma ^{-2j-k} e^{-\sqrt{\frac {2}{k-2}} (j+\frac k2)\phi } e^{i\sqrt{\frac 2k} (j+\frac k2) X} e^{i\sqrt{\frac 2k} (j+\frac k2 \omega) T} :  \times  \ h.c.  \label{martinazo}
\end{equation}
This represents an alternative form for the highest-weight state $\Phi _{j,j,j} ^{\omega }$ and should be incorporated to the {\it bestiary} studied in \cite{nos2}. This operator includes the known \cite{fzz,mo3} conjugate representation for the identity 
\begin{eqnarray}
\baseRing{I} \sim \hat {\Phi }_{0,0,0}^{0} = : \gamma ^{-k} e^{-\frac{k}{\sqrt{2k-4}} \phi } e^{i\sqrt{\frac k2} X} : \times \ h.c. \ \ , \ \ \hat {\Phi }_{-\frac k2,-\frac k2,-\frac k2}^{1} = \Phi ^0 _{0,0,0} = 1 
\end{eqnarray}
And the following relations hold in general 
\begin{eqnarray}
\hat {\Phi }_{-\frac k2-j,-\frac k2-j,-\frac k2-j}^{\omega +1} = {\Phi }_{j,-j,-j}^{\omega } \ , \ \ \ \ \tilde {\Phi }_ {-\frac k2 -j,\frac k2 +j,\frac k2 +j}^{\omega -1} = {\Phi }_{j,j,j}^{\omega } . \label{catorce}
\end{eqnarray}
The main observation leading to (\ref{catorce}) relies in the fact that the twisting of the $U(1)$ component ({\it i.e.} the difference $ k\omega /2$ between the charges of both free fields $X$ and $T$) and the difference of the signatures of both auxiliary bosons permit to build up the $\omega $inding $\omega$-sectors of $SL(2,\R )_k /U(1) \times U(1)$ by assembling these on the sectors $\omega \in \{ -1,0,+1\}$ of the $SL(2,\R )_k$ factor. Then, the identification between discretes representations of index $j$ and $-\frac k2 -j$ turns out to be rigidly translated to generic $\omega $-sectors.
\subsection{A remark on screening operators and conjugate representations}

Conjugate representation (\ref{conj+-}) also permits us to write a combined interaction term $S_I+\tilde S_I$ composed by (\ref{arriba}) and (\ref{arribasss}) in the following form\footnote{Let us notice that this fact does not mean that the interaction term can be reproduced by using the Wakimoto representation (\ref{asdf1}) in similar way, as it can be seen from the operator product expansion.}
\begin{eqnarray}
S_{I}+\tilde S _{I}= \frac {M}{4\pi }\int d^2z \left( \tilde {\Phi}^{-1 }_{1-\frac k2, \frac k2,\frac k2} - \Gamma (k-1) f_{M } (k-2) \tilde {\Phi}^{-1 }_{\frac k2 -2,\frac k2,\frac k2}  \right)  \label{arribalatinos}
\end{eqnarray}
where $f_{M } (k-2) = f_1 (k-2) M ^{k-3}$. It is reminiscent of the case of Liouville theory 
\cite{teschnerliouville}, where the (auto)interaction term can be written as a particular evaluation of the functional form of the vertex operators. 

In order to extend the analogies with the Liouville model, let us notice that by renormalizing the black hole mass as $M \rightarrow M _{reg}=M \frac {\Gamma (\frac {1}{k-2})}{\Gamma (1- \frac {1}{k-2})}$ and taking the limit $k \rightarrow 3$, the following expression for the (composed) interaction term is obtained
\begin{eqnarray}
S_{I}+\tilde S_{I}=\frac {M _{reg}}{4\pi }\int d^2z \beta \bar \beta e^{-\sqrt {2} \phi } \left( \sqrt {2} \phi -\log \pi M _{reg} \beta \bar \beta  \right) \label{arribasudacas}
\end{eqnarray}
which is analogous to the $c=25$ model, where it has been proved that the Liouville interaction term receives similar corrections \cite{teschnerliouville} (see also \cite{estospibes, ultimo, conclaudio}). In the $D=1$ non-critical string theory it was argued that these non-exponential operators control the anomalous scaling behaviour of random surfaces \cite{jp}; thus, the analogous critical behaviour in the $k=3$ limit of non-compact WZNW model turns out to be an interesting aspect. 

Actually, the critical point $k=3$ deserves attention because of the appearence of particular properties there; {\it e.g.} it can be proven that in such case the unitarity bound on the free spectrum implies the locality of the dual conformal field theory formulated on the boundary of $AdS_3$ space, without the requirement of additional constraints on the external states in the $N$-point correlators. 

On the other hand, notice that the appearence of a linear term in $\phi $ and a logarithmic term in $\beta \bar \beta$ in expression (\ref{arribasudacas}) is not a surprise since it is precisely consistent with the fixing $\beta \bar \beta \sim \mu _L$ when analysing, for example, the correspondence\footnote{since a logarithmic dependence of the Liouville cosmological constant $\sim \log \mu _L$ arises in the resonant point $2\alpha =b+b^{-1}$ of the Liouville reflection coefficient $R(\alpha )$.} between the $c=1$ David-Distler-Kawai theory and the special superconformal coset $\hat c =3$ studied in \cite{vafa} in the context of the Knizhnik-Polyakov-Zamolodchikov formalism. 

The ingredients of this discussion are also reminiscent of the Liouville reduction of WZNW action. Actually, let us briefly remind that by expanding the WZNW theory arround a classical solution satisfying $J^+=k$, $J^-=J^3=0$, the Sugawara stress-tensor yielding from (\ref{laconc})-(\ref{hadetumadre}) takes a form which, without difficulties, is recognized as the one corresponding to the Liouville stress-tensor, where the Liouville field $\varphi $, the cosmological constant $\mu _L$ and the background charge $Q_L $ are given by
\begin{eqnarray*}
\phi = - \sqrt {2} \varphi ,  \ \ \mu _L = k M _{reg} , \ \ Q_L = b+b^{-1} , \ \ b^{-2}=k-2
\end{eqnarray*}
Moreover, (\ref{arriba}) and (\ref{arribasss}) are then identified with the screenings of Liouville theory. The physical interpretation and the geometrical meaning of this reduction were explained\footnote{The configuration $J^+=k$ and the Liouville reduction were studied in detail in references \cite{sw,br,botros}. We will not discuss here the issues of the {\it twisting} of stress-tensor ${\mathcal T} \rightarrow {\mathcal T}+ \partial J^3$, the BRST cohomology of additional discrete states, the description of the single long string configuration and the relaxation of the constraint $J^3=0$; we refer to the mentioned references for the details about these aspects.} in \cite{sw,br}. Then, we see from (\ref{arribasudacas}) how the fixing $\beta \bar \beta \sim \mu _{L}$ survives in the limit $b \rightarrow 1$.


Then, as an additional corollary, we emphasize that the critical points $k=3$ and $k=1$ of the $SL(2,\R )_k$ WZNW model, which are the fixed points of the transformation $(k-2) \rightarrow (k-2)^{-1}$, seem to be an interesting subject for further study; in particular, its connection with the $c=1$ model.

The class of operators (\ref{arribasudacas}) was also studied in reference \cite{gastonlog} within the context of the prelogarithmic representation $j=-\frac 12$ of $\hat {sl(2)} _k$ algebra. 

\subsection{Two-point correlation function}

Now, let us consider the two-point correlation functions, which will be denoted ${\mathcal A}  ^{j_1,j_2}_{m_1,m_2} = < {\Phi}^{\omega _1}_{j_1,m_1,\bar {m} _1}(z_1){\tilde {\Phi}}^{\omega _2}_{j_2,m_2,\bar {m} _2}(z_2)>$. A direct calculation of two and three-point functions in terms of Dotsenko-Fateev type integrals on the whole complex plane can be performed with the purpose of obtaining the explicit value for $B(j)$ in (\ref{efffe}), being the two-point correlator of the form (see \cite{teschner})
\begin{equation}
{\mathcal A} ^{j_1,j_2,j_3}_{m_1,m_2} =\left| z_1-z_2\right| ^{-4h _{(j_1,m_1)}}\left( R (j_1,m_1)\delta (j_1-j_2)+\delta
(j_1+j_2+1)\right) \delta (m_1+m_2)
\end{equation}
where
\begin{equation}
B(j) = (2j+1) \left( - \frac{\pi M \Gamma (\frac 1{k-2})}{\Gamma (1-\frac 1{k-2})%
}\right) ^{2j+1}\frac {\Gamma \left( 1-\frac{2j+1}{k-2}\right)}{\Gamma \left( \frac{2j+1}{k-2}\right)}  \label{herre}
\end{equation}
The {\it additional} overall factor $2j+1$ standing in (\ref{herre}) was obtained in reference \cite{becker}; its presence was remarked in \cite{nos3} and explained in \cite{mo3}, where it was pointed out that it yields from the direct evaluation of the delta factor $\delta(j_1-j_2)$ in the two-point function and because of the projective invariance codified in the presence of the factor $Vol ^{-1}(PSL(2,\C))$. In fact, this factor signals the differences existing between continuous and discrete representations of $SL(2,\R )$. 

The steps of the calculation of expression (\ref{herre}) are analogous to those followed in \cite{nos3}, {\it i.e.} the expression
\begin{equation}
\lim_{\varepsilon \rightarrow 0}\frac{\Gamma (\varepsilon )}{\Gamma (\varepsilon -s)}%
=(-1)^s\Gamma (s+1)  \label{trucus}
\end{equation}
(for $s\in \N$) needs to be considered and combined with the charge symmetry conditions
\begin{eqnarray}
j_1-j_2= s \ , \ \ m_1+m_2= 0  \ , \ \ \omega _1+\omega _2= 0  \label{nner}
\end{eqnarray}
in order to turn the factor
\begin{eqnarray}
 \frac {\Gamma (s-j_1+j_2+\bar m _1+\bar m _2)}{\Gamma (-j_1+\bar m _1)\Gamma (-j_2+\bar m _2)} \frac {\Gamma (s-j_1+j_2+m_1+m_2)}{\Gamma (-j_1+m_1)\Gamma (-j_2+m_2)} ,  \label{truco}
\end{eqnarray}
coming from the multiplicity of the Wick contractions of the $(\beta ,\gamma)$ system, into
\begin{eqnarray}
\frac {\Gamma (j_1-m_1+1)\Gamma (j_2-m_2+1)}{\Gamma (\bar {m}_1-j_1)\Gamma (\bar {m}_2-j_2) } .
\end{eqnarray}
In (\ref{nner}) and (\ref{truco}), the number $s$ refers to the total amount of screening operators that needs to be included in order to render the correlation functions non vanishing. 
\subsection{Dotsenko prescription and path integral approach}

In the consideration above, we used conditions (\ref{nner}), which are defined by the specific conjugate representation of the identity operator
\begin{equation}
\baseRing {I} \sim \tilde{\Phi}^{0}_{0,0,0} = : e^{-\frac{k}{\sqrt {2k-4}}\phi } e^{-i\sqrt{\frac {k}{2}}X}  : \times  \ h.c. \label{estoyaca1}
\end{equation}
The derivation of these conservation laws was detailed in \cite{nos2} and \cite{dot1}. These conservations laws are such that the power of $\gamma $ fields and the power of $\beta $ fields do coincide for non-vanishing correlators. For instance, this fact enabled the authors of \cite{petko} to discuss in detail the correlators containing rational powers of these commuting fields. 


\subsubsection*{Dotsenko prescription}

The conjugate representations were originally studied by Dotsenko in the context of $SU(2)_k$ WZNW model in \cite{dot1}; and were analyzed in \cite{nos2} for the 2D black hole geometry. Basically, the construction includes additional elements, namely the conjugate (alternative) representations of vertex operators in the theory. The underlying idea is that each given conjugate representation (let us say $\tilde {\Phi } _{j,m,\bar m}^{\omega }$) induces a non-trivial representation of the identity operator, which is given by a zero-dimension field corresponding to the evaluation $\tilde {\Phi} _{0,0,0}^{0} (z) \sim \baseRing{I}$. Then, the balance of the {\it charges} of this {\it identity operator} under the different fields is translated into particular charge symmetry conditions for the states involved in a non-vanishing correlation function (see the original work \cite{dot1}). Eventually, this conditions turn out to be conservation laws for non-vanishing scattering amplitudes.


Because of the fact that this construction can seem to be a little abstract up to this point, let us clarify the idea by giving a concise example: Notice that, actually, (\ref{estoyaca1}) is not the unique {\it conjugate identity}; in fact, other zero-dimension operators can be used as a conjugate representation of the identity, $v.g.$ let us consider
\begin{equation}
\baseRing {I} \sim \Phi^{0}_{-1,0,0} =  : \gamma ^{-1}e^{-\sqrt {\frac{2}{k-2}}\phi } :  \times  \ h.c. \label{p82}
\end{equation}
The consideration of this operator as the conjugate representation of the identity leads to certain charge symmetry conditions (in the spirit of \cite{dot1}) which precisely coincide with the conservation laws yielding from the direct integration over the zero-mode\footnote{On the other hand, the consideration of the conjugate identity $\tilde {\Phi} ^0 _{0,0,0}$ leads to conservation laws (\ref{nner}).} $\phi _0$ in the action $S_1$. This is because the background charge $-\sqrt {\frac {2}{k-2}}$ of the linear dilaton term in WZNW action coincides with the {\it charge} of the conjugate representation (\ref{p82}) under the field $\phi $. Moreover, the power $-1$ of the field $\gamma $ in (\ref{p82}) coincides with the difference of powers of fields $\gamma $ and $\beta $ required in order to have non-vanishing correlators, as it is determined by the Riemann-Roch theorem on the sphere. Besides, the zero-dimension operator (\ref{p82}) induces a proper conjugate representation for the field (\ref{asdf1}); and this conjugate version of the field $\Phi ^{\omega }_{j,m,\bar m} $ is simply the {\it Weyl reflected} operator 
\[
\bar {\Phi } ^{\omega }_{j,m,\bar m} = \Phi ^{\omega }_{-1-j,m,\bar m} 
\]
and, consequently, we could write $\bar {\Phi } ^{0}_{0,0,0}$=(\ref{p82}).

In this sense, operator (\ref{p82}) could be interpreted as the {\it background charge} operator.

Hence, this discussion suggestes to try to understand the standard Coulomb gas prescription \cite{becker} as a particular case which can be included within the framework of this extended Dotsenko-like realization. On the other hand, this permits us to suggest a connection with the arising of zero-dimension operators in the expression for three-point funcions in previous computations, see \cite{satohf}. Certainly, operator (\ref{p82}) can be associated to the leading term in the large $\phi$ regime of certain mode $(i.e.$ $j+1 = m =\bar m = 0) $ of the operator
\begin{eqnarray*}
\lim _{j\to -1 } \sum _{n=0}^{\infty } \frac {(-1)^n e^{-\sqrt {\frac{2}{k-2}} (n+1)\phi }}{\Gamma (n+1) \Gamma (n+2)} \int d^2 x \ x^{j-m} \bar x ^{j-\bar m} \frac {\partial ^n}{\partial \gamma ^n} \delta (\gamma -x) \frac {\partial ^n}{\partial \bar \gamma ^n} \delta (\bar \gamma - \bar x) 
\end{eqnarray*}
which is the non-trivial spin $j=-1$ primary field which emerges in the path integral approach to string theory on $AdS_3$. The insertion of this field was considered in order to define the path integral representation of correlation functions\footnote{notice that the substitution $j \rightarrow -1-j$ is required when comparing with references \cite{satohf}.}, see \cite{satohf}.



\subsection{Three-point function violating winding number}

Now, let us move to the three-point functions ${\mathcal A} ^{j_1,j_2,j_3}_{m_1,m_2,m_3} $. Since our principal purpose is to study the three-point correlators describing interaction processes violating the $\omega $inding number conservation, let us discuss the general form of these non-conservative cases in particular. As mentioned, an explicit formula ({\it cf.} \cite{mo3}) for this kind of quantities was given in reference \cite{nos3} for particular cases with two states being of highest or lowest-weight, {$e.g.$ $j_2=-m_2$ and $j_3=-m_3$}. Here, we can generalize that formula by including the operator (\ref{conj+-}) in order to consider a generic value for $j_2$ and $m_2$. Thus, by using (\ref{trucus}) and the fact that the following charge symmetry conditions hold
\begin{eqnarray}
j_1 -j_2-j_3+ \frac {k}{2} = s \ , \ \ m_1+m_2+m_3-\frac k2 = 0 \ , \ \ \omega _1+\omega _2 +\omega _3 +1 = 0,  \label{wina}
\end{eqnarray}
we find that the three-point functions violating the $\omega $inding number conservation in one unit is proportional to
\begin{eqnarray}
(-1)^{m_2-\bar {m}_2}\frac {\Gamma (j_1-m_1+1)\Gamma (j_2-m_2+1)}{\Gamma (-j_1+\bar {m}_1)\Gamma (-j_2+\bar {m}_2)}  \label{trucoa}
\end{eqnarray}
which is the group theoretical factor ({\it cf.} \cite{fh}). In order to write the multiplicity factor yielding from the Wick contraction of the $(\beta ,\gamma )$ system in a simple way, we find convenient to consider the generic value of $m_2$ but still keeping the lowest-weight condition for the third operator, {\it i.e.} $m_3+j_3=0$. In fact, the complete expression for the three-point function ${\mathcal A}  ^{j_1,j_2,j_3}_{m_1,m_2,-j_3} = <{\Phi}^{\omega _1}_{j_1,m_1, \bar {m} _1}(0){\tilde {\Phi}}^{\omega _2}_{j_2,m_2,\bar {m} _2}(1) {\tilde {\Phi}}^{\omega _3}_{j_3,-j_3,-j_3}(\infty )>$ in such case turns out to be 
\begin{eqnarray}
&&{\mathcal A}  ^{j_1,j_2,j_3}_{m_1,m_2,-j_3} =(-1)^{m_2-\bar {m}_2} \left( - \frac { \pi M \Gamma (\frac {1}{k-2})}{\Gamma (1-\frac {1}{k-2})}  \right)^s  \frac {\Gamma (j_1-m_1+1)\Gamma (j_2-m_2+1)}{\Gamma (-j_1+\bar {m}_1)\Gamma (-j_2+\bar {m}_2)} \times \nonumber  \\
&&\times \frac{ G_{k}\left( -1-\sum_{a=1}^{3}j_{a}-\frac{k}{2}\right) G_{k}\left(
-j_{12}-\frac{k}{2}\right) G_{k}\left( 1+j_{13}-\frac{k}{2}\right)
G_{k}\left( -j_{23}-\frac{k}{2}\right) }{G_{k}\left( -1\right) G_{k}\left(
-2j_{1}-1\right) G_{k}\left( -2j_{2}-k\right) G_{k}\left( 2j_{3}+1 \right) } \nonumber \\
\label{ias}
\end{eqnarray}
where $j_{ab} \equiv 2j_{a}+2j_{b}-\sum _{c=1} ^3 j_{c}$ and where the $G_{k}$ functions are given in terms of the double Barnes functions $\Gamma _{2}$ by the following expression 
\[
G_{k}(x) = (k-2)^{\frac{x(k-1-x)}{2k-4}}\Gamma _{2}(-x, 1,k-2)\Gamma _{2}(k-1+x, 1,k-2) 
\]
being 
\[
\log (\Gamma _{2}(x, 1,y)) = \lim_{\varepsilon \rightarrow 0}\frac {\partial}{
\partial \varepsilon } \left( \sum_{n=0}^{\infty }\sum_{m=0}^{\infty
}(x+n+my)^{-\varepsilon } - \sum_{n=0}^{\infty }\sum_{m=0}^{\infty
}(1-\delta _{n,0}\delta _{m,0}) (n+my)^{-\varepsilon }\right)
\]
The expression (\ref{ias}) is proportional to a factor $\delta(\omega _1+\omega _2 +\omega _3 +1)$ which stands from the integration over the free fields $X$ and $T$. This expression basically corresponds to the standard three-point function by replacing the quantum number as $j_2 \rightarrow -j_2-\frac k2$, $j_3 \rightarrow -1-j_3$. Moreover, since the affine properties of conjugate representations are defined up to a $m$-independent factor, then the considerations regarding the relative normalization between different representations also hold in this analysis. For instance, the formula (\ref{ias}) can receive an additional factor $\sim B(j_2)$ given by the normalization of the second vertex $\tilde {\Phi } ^{\omega _2}_{j_2,m_2,\bar {m} _2}$.

A fact which will turn out to be crucial in the last section is that the $m$-independent pole conditions of the formula (\ref{ias}) are located at
\begin{eqnarray}
1+j_{1}+j_{2}-j_{3}-\frac{k}{2} &\in &\Z_{\geq 0}+\Z_{\geq
0}(k-2) \ , \ -j_{1}-j_{2}+j_{3}-\frac{k}{2} \in \Z_{\geq 0}+\Z_{\geq 0}(k-2) \label{b2}\\
1-j_{1}+j_{2}+j_{3}-\frac{k}{2} &\in &\Z_{\geq 0}+\Z_{\geq 0}(k-2) \ , \ j_{1}-j_{2}-j_{3}-\frac{k}{2} \in \Z_{\geq 0}+\Z_{\geq
0}(k-2) \label{b1} \\
1+j_{1}-j_{2}+j_{3}-\frac{k}{2} &\in &\Z_{\geq 0}+\Z_{\geq
0}(k-2) \ , \ -j_{1}+j_{2}-j_{3}-\frac{k}{2} \in \Z_{\geq 0}+\Z_{\geq 0}(k-2) \label{b6} \\ -1-j_{1}-j_{2}-j_{3}-\frac{k}{2} &\in &\Z_{\geq 0}+\Z_{\geq 0}(k-2) \ , \ 2+j_{1}+j_{2}+j_{3}-\frac{k}{2}\in \Z_{\geq 0}+\Z_{\geq 0}(k-2) \label{b8}
\end{eqnarray}
Within the context of the $SL(2,\R )_k /U(1) \times U(1)$ WZNW model, the divergences (\ref{b2})-(\ref{b8}) can be undertood in terms of the analysis presented in \cite{mo3}: these come from the integration over the spacetime coordinate $\phi $; and the shifting in $-\frac k2$ precisely corresponds to the fact that the string states have no defined $\omega $ number and in the case of $\omega $inding violating processes the worldsheet operator mainly contributing to the pole takes the form $\sim e^{\sqrt{\frac {2}{k-2}}(-j-\frac k2) \phi}$ instead $\sim e^{\sqrt{\frac {2}{k-2}}j \phi}$ (see (\ref{conj+-}) and (\ref{martinazo}), {\it cf.} (\ref{asdf1})).

The formula (\ref{ias}) is (up to a minor generalization) the free field computation of correlators describing the scattering processes violating the $\omega $inding number conservation in string theory in $AdS_3$ presented in \cite{nos3}. However, we obtained here the factor (\ref{trucoa}) which, among other things, enables us to perform a comparison with (its dual) sine-Liouville field theory. We will show that, in particular, such $m$-dependent group theoretical factor is in good correspondence with the one arising in the dual model. Besides, we can also notice that similar factor was found in reference \cite{mo3} for string scattering processes violating the $\omega $inding number conservation in $AdS_3$ geometry, where the computation includes the coincidence limit of the {\it spectral flow operator} and one of the vertex involved in the interaction. By using (\ref{trucus}), it is feasible\footnote{for this purpose, it is necessary to rename the index of the third vertex as $4 \rightarrow 3$ and conciliate the nomenclatures by changing $j \rightarrow j+1$. In order to perform a comparison with the formula presented in \cite{mo3}. It could be also helpful to take into account that the following equation holds $G_k (-1-j_1-j_2-j_3-\frac k2)=(k-2)^{2(j_1+j_2+j_3+1)+3-k} \frac {\Gamma (4+j_1+j_2+j_3-\frac k2)}{\Gamma (-3-j_1-j_2-j_3+\frac k2)} G_k (-3-j_1-j_2-j_3+\frac k2)$; this permits to show the exact agreement existing between the analytic properties of the formulas for three-point functions presented in \cite{nos3} and \cite{mo3}.} to write the factor standing in the expression found in \cite{mo3} as $(-1)^{\bar {m} _3-m_3} \prod _{a=1}^{3} \frac {\Gamma (1+j_a -m_a)}{\Gamma (-j_a +\bar {m} _a)}$. 

As we just mentioned, as a consistency check of the formula for $\omega $inding violating three-point function in WZNW theory, we can perform a comparison with the analogous result known for sine-Liouville dual theory. The residues for correlators of winding violating processes in sine-Liouville theory were computed in \cite{fh}. Indeed, a careful computation shows that the analytic structure of such observables in both CFT's exactly agree. We give the details of the comparison in section 5.

\section{Remarks}

The free field description of non-compact two-dimensional conformal
field theories turned out to be a useful tool to work out the features of
this class of non-trivial models. In \cite{nos3,becker,satohf}, among many other papers, the scope of
this representation proved to be suitable to obtain the whole expression for
the two and three-point functions in $SL(2,\R)_k$ WNZW model. In this note, our intention was to revisit and extend its study.


\subsubsection*{{\it Addemdum} to references \cite{nos2} and \cite{nos3}}

We presented here a continuation of the study of references \cite{nos2,nos3} regarding the Dotsenko conjugate representations of $\hat {sl(2)} _k$ vertex algebra. We pointed out the relevance of operators (\ref{conj+-})-(\ref{arribasudacas}) in the context of free field representations; and we remark that these should be included in the {\it bestiary} in order to observe interesting features of the construction and consider the discussion complete. For example, we discussed the relation between the conservation laws yielding from the integration over the zero-mode of free fields and the Dotsenko prescription for the charge symmetry conditions (see (\ref{p82})). We also discussed the $k \rightarrow 3$ limit of the screening operators, whose functional form (\ref{arribasudacas}) manifestly shows the connection with the $c=1$ matter model, and we extended the Dotsenko-like highest-weight operators \cite{nos2,nos3} belonging to the conjugate representations of $SL(2,\R )$. 

We showed how to construct $\beta$-dependent conjugate representations of vertex operators describing $\omega $inding states in $SL(2,\R )_k /U(1)\times U(1)$ by extending the free field representation of the simplest discrete states on the two-dimensional black hole; obtaining, in this way, Kac-Moody primary states on the product $SL(2,\R )_k/U(1) \times U(1)$ (see (\ref{conj+-}))\footnote{Besides, other operators refered here, $e.g.$ (\ref{estoyaca1}), were previously studied in the literature; for very interesting studies of world-sheet operators see \cite{ads3,fzz,mo,mo3,dot1,giveon,dot2}.}.

\subsubsection*{A word on supersymmetry}

As we mentioned in the introduction, the conformal model formulated on the product $SL(2,\R )/U(1) \times U(1)$ appears in different contexts, {\it v.g.} the proof on the non-ghost theorems on $AdS_3$ and the constructions of models with N=2 supersymmetries. For instance, in relation to this aspect, in \cite{susy1} it was shown how to construct this class of supersymmetric models on the product $AdS_3 \times {\mathcal N}$, being ${\mathcal N}$ a compact manifold containing a $U(1)$ affine symmetry. The construction basically follows the steps of Kazama-Suzuki method; {\it i.e.} three free fermions $\psi ^a \ a \in \{+,-,3\}$ are considered in order to realize the superconformal extension of the $\hat{sl(2)}_k$ algebra, where the Kac-Moody level is shifted as usual. An additional free boson and a free fermion representing the $U(1)$ of ${\mathcal N}$ have to be included as well. In reference \cite{susy2} the authors explained in detail how to split the $U(1)$ component of the factor ${\mathcal N}$ in order to bosonize the fermions realizing the $\hat {sl(2)} _k$ algebra. The important point which makes contact with our discussion is the fact that the conjugate representations we were describing here can be systematically extended to the case with N=2 worldsheet supersymmetries in a straightforward way. For example, it is feasible to see that non-trivial zero-weight operators (perhaps describing conjugate representations of the identity operator) can be constructed by a (non-unique) linear combination associating the {\it identities} $\Phi  ^{\pm 1} _{-\frac k2,\mp \frac k2,\mp \frac k2}$ and the free fermions $\psi ^{\pm}$, once the shifting $k \rightarrow k+2$ is adequately taken into account. The $\omega $inding violating processes were discussed in \cite{gk2} within the context of the N=2 supersymmetric version of the theory in a rather different approach, which involves the twisting of the supercharges which have to be inserted in the correlators to change the picture of the operators. Here, we restricted our discussion to the bosonic theory. The supersymmetric theory is studied in the recent paper \cite{diego}.

\subsubsection*{Conjugate representations in $SL(2,\R )_k /U(1)\times U(1)$ WZNW theory}

A deeper understanding of the connection between the discrete states in two-dimensional black hole geometry and the conjugate representations we presented here is, of course, desirable. Here, we have found interesting pointing out its existence, to suggest the mentioned connection and to remark the main properties.

Then, let us summarize what, in our opinion, are the principal properties of the conjugate representations $\tilde {\Phi } ^{\omega }_{j,m,\bar m}$. We do this as follows: 

The first important thing is that representation $\tilde \Phi _{j,m,\bar m}^{\omega }$ includes the Dotsenko conjugate representation \cite{dot3} for the particular case $\tilde {\Phi }^{-1} _{j+\frac k2-1,j+\frac k2,j+\frac k2}$. And it also generalizes the highest-weight states $\tilde \Phi _{j,j,j}^{0}$ considered in reference \cite{petko2}. The $\beta$-dependent representation presented here also generalizes the funcional forms recently proposed in \cite{diego} (see eq. (2.101) of this reference). On the other hand, this representation includes the free field representation of Bershadsky-Kutasov discrete states \cite{bk} on 2D black hole for the particular evaluation $\tilde {\Phi }^{-1} _{-\frac k2-j,m+\frac k2,\bar m +\frac k2}$ (see \cite{discretos} and references therein). And, precisely because of this, we observe that both screening operators ${\mathcal L} _I$ and $({\mathcal L} _I)^{k-2}$ can also be written as particular vales, namely $\tilde \Phi _{1-\frac k2,\frac k2,\frac k2}^{-1}$ and $\tilde \Phi _{\frac k2 -2,\frac k2,\frac k2}^{-1}$ respectively. 


The representation $\tilde \Phi _{j,m,\bar m}^{\omega }$ has the same conformal properties (under Kac-Moody and Virasoro algebras) of the representation $\Phi _{j,m,\bar m} ^{\omega }$. In this sense, we achieved to find a free field realization which is equivalent to $\Phi _{j,m,\bar m}^{\omega }$ but presents a functional form which goes like $\sim e^{\sqrt {\frac {2}{k-2}}(-j-\frac k2) \phi}$ (instead $\sim e^{\sqrt {\frac {2}{k-2}}j \phi}$) in the large $\phi $ limit. This permits us to understand, for example, the divergences of the three-point functions (\ref{ias}) in terms of the integration over the target space coordinate $\phi $. 

In relation with this last point, we find that when representations $\tilde \Phi _{j,m,\bar m}^{\omega }$ are used (instead the standard Wakimoto form $\Phi ^{\omega } _{j,m,\bar m}$) to compute correlation functions, the conservation laws coming from the integration over the zero-mode leads to violate the ${\omega }$inding number. It is also closely related to the fact that certain identities satisfied by the evaluations $\hat {\Phi} ^{\omega -1}_{j,j,j}$ and $ {\Phi} ^{\omega +1}_{j,-j,-j} $ and $\Phi ^{\omega }_{-\frac k2-j, \pm \frac k2 \pm j, \pm \frac k2 \pm j}$ realize the spectral flow symmetry of $\hat {sl(2)}_k$ affine algebra.

The last properties enumerated above suggest an interpretation for representation $ \tilde {\Phi} ^{\omega }_{j,m,\bar m}$ as the one describing the $\omega =1$ contribution of the wave function of a string state in $AdS_3$. {\it i.e.} as it was pointed out in \cite{mo3}, the states in $AdS_3$ have no defined $\omega $ number and then, even for states with well defined energy, one should write the corresponding wave function as a linear combination of different contributions \cite{mo3}. This manifests the fact that $\omega $inding number is not a conserved quantity in this space. Schematically, this is represented by the series
\[
\Psi _{j,m,\bar m} = \Psi ^{(\omega =0)}_{j,m,\bar m} + \Psi ^{(\omega =1)}_{j,m,\bar m} + ...
\] 
Consequently, we are claiming that the following realization holds for this
\[
\Psi _{j,m,\bar m} = \Phi ^{0}_{j,m,\bar m} + \tilde {\Phi } ^{0}_{j,m,\bar m} + ...
\]
where we are proposing this meaning for the conjugate representation $\tilde {\Phi } ^{\omega =0} _{j,m,\bar m}$ we have written down. Notice that it is not simply the sum of states with different $\omega $ numbers.

In this context, we used the free field representation $ \tilde {\Phi} ^{\omega }_{j,m,\bar m}$ to write the formula (\ref{ias}) for correlators describing processes which violate the $\omega $inding number conservation in $SL(2,\R )_k/U(1) \times U(1)$ WZNW model. And we obtained the ($m$-dependent) group theoretical factor (\ref{trucoa}).

\subsubsection*{Correlation functions and duality}

Once the construction of analogous correlation functions representing
processes which violate the total $\omega $inding number in sine-Liouville conformal
field theory is analysed within the context of the prescription
proposed in \cite{fh}\footnote{we describe it in section 5.}, we are able to perform a comparison of
the analytic structure of this class of (non-conservative) three-point correlators for both conformal models. 

In previous works on these subjects, the discussion about the integration over the zero-mode of sine-Liouville fields turned out to be a relevant point when studying the analytic structure of two-point \cite{fzz,kkk} and three-point \cite{fh} $\omega $inding conserving processes. We observe here that the $\omega $inding violating processes are not the exception, since the analysis of these observables leads us to obtain the correct overall factor standing in the correlators which precisely yields from the integration over the zero-mode in sine-Liouville CFT. 

This consideration is an important step because if one naively compare the formula for the residue of sine-Liouville correlators presented in reference \cite{fh} with the three-point functions computed in \cite{nos3,mo3} for WZNW theory (without taking into account the divergent factor yielding from the integration over the zero-modes) the pole structures of both models would not agree. Thus, we write the complete analytic structure for sine-Liouville three-point functions in the netx section in order to compare the observables of these CTF's properly.
 
Then, we are able to show that the analytic structure of both theories are in good agreement; including the ($m$-dependent) group theoretical factor, the location of the zeros, the phase and the order and positions of the poles (see section 5). This
result extends the previous analysis of reference \cite{fh} since includes the $\omega $inding violating processes in the framework of the comparison. On the other hand, this presents another result in favor of checking particular cases of the
FZZ conjecture, which, as mentioned, states the duality
between the sine-Liouville model and the string theory on the euclidean
version of two-dimensional Witten black hole, {\it i.e.} the cigar manifold. Similar results were attributed to the seminal unpublished work \cite{fzz}.


\section{On Three-point function in sine-Liouville field theory}

In a more general context, several points could be mentioned as motivations to study the problem of two
dimensional string theory in a $\omega $inding background. An important example of this kind
of conformal models is the sine-Liouville string theory, which has been
recently discussed as an example of non trivial CFT belonging to a more
general family of possible backgrounds of two-dimensional string theory \cite
{akk,kostov2}; it was also studied in relation with the description of
strong vortex and tachyon perturbations \cite{kostov}.

The purpose of this section is to study the correlation functions violating $\omega $inding number in sine-Liouville CFT. To me more precise, we will perform a comparison between these correlators and the analogous computation in WZNW we described in section 3. 

In fact, if one assumes the duality existing between these CFT's, then this comparison represents a non-trivial check of the formulas for the observables computed in both models. This is, certainly, the main goal of this section.

\subsubsection*{On Fateev-Zamolodchikov-Zamolodchikov conjecture}

One of the most interesting results which can be found among the
recent subjects of research is, perhaps, the conjectured strong-weak coupling
duality relating the sine-Liouville conformal field theory and the
WZNW model formulated on $SL(2,\R)/U(1)$ group manifold.
This celebrated result is known as the Fateev-Zamolodchikov-Zamolodchikov
conjecture (FZZ) \cite{fzz} and, in the last years, it was successfully used
in the study of properties of degenerated operators in non-compact conformal
theories \cite{giveon}, in the computation of correlation functions in
string theory on euclidean $AdS_{3}$ and in the study of quantum corrections
of the entropy of two-dimensional black hole \cite{kkk} (see also the recent 
papers \cite{i912} and \cite{pendejo}).

It is usually suggested that the FZZ
conjecture could be derived as some class of perturbation or reduction of other known
results related to mirror symmetry \cite{hori2}; indeed, a supersymmetric version of
the duality has been proved, which states the equivalence of the
$SL(2,\R)/U(1)$ Kazama-Suzuki model and N=2 super Liouville theory \cite{hori}.
With the purpose to learn about other interesting results related to
N=2 Liouville theory we draw the atenttion to reference \cite{russo}.

Here, we address our attention to the analysis of correlation functions in sine-Liouville CFT.

In this context, let us essay a comparison between the analytic properties of the $\omega $inding violating three-point correlators in WZNW model with similar computations made in the sine-Liouville model. Our input will be the results presented in the previous subsection and in reference \cite{fh}, which contain the formulas for three-point functions describing processes violating the $\omega $inding number conservation in both CFT's respectively.

Comparisons between correlators of both models were previously considered, but restricted to the $\omega $inding conserving processes; we present here the case of $\omega $inding violating interactions. Similar analysis was attributed to the unpublished paper \cite{fzz}.

\subsubsection*{Sine-Liouville conformal field theory}

Within the context of the geometrical interpretation of Liouville model as a
two-dimensional string theory, the sine-Liouville theory represents the
condensation of vortices. The action of this theory is given by 
\begin{equation}
S_2=\frac{1}{4\pi }\int d^{2}z\left( \partial \phi \bar {\partial }\phi -\sqrt {\frac {2}{k-2}} R \phi +\partial X \bar {\partial }X+ \lambda
e^{-\sqrt{\frac{k-2}{2}}\phi }\cos \left( \sqrt{\frac{k}{2}}\left( X
_{L}-X _{R}\right) \right) \right)   \label{h22}
\end{equation}
where $X(z,\bar z)= X_R(z)+X_L(\bar z)$.

Fukuda and Hosomichi proposed in \cite{fh}  a free field realization of the correlation functions in this
theory by means of the insertion of two different screening operators%
\footnote{The screening operators realize the interaction term of sine-Liouville
action; see (\ref{congo}).} and by evaluating a generalized Dotsenko-Fateev type
integrals. The three-point correlators representing processes
violating $\omega $inding number were explicitly computed by the authors; this was
achieved by the insertion of a different amounts of both screening charges.

In the context of formal aspects, similar integral representations have
been discussed, for instance, in the free field realization of string theory on euclidean
$AdS_{3}$ and in the case of two-dimensional black hole background (see \cite{becker,satohf} and references therein). However, it is important to mention that several
differences can be noticed if a detailed comparison of these models is
performed; the role played by the screening charges and the conjugate
representations of the identity operator are examples of these distinctions.

The vertex operators creating primary states from the vacuum of the theory
can be written in terms of the bosonic free fields $\left( \phi ,X
\right) $ as 
\begin{equation}
\Phi _{j,m,\bar{m}}(z,\bar{z})= : e^{j\sqrt{\frac{2}{k-2}}\phi (z,\bar{z}%
)+i\sqrt{\frac{2}{k}}\left( mX _{L}(z)+\bar{m}X _{R}(\bar{z}%
)\right) } :  \label{fuld}
\end{equation}
The conformal dimensions of such operators are given by\footnote{Indeed, the nomenclature here is usual in the treatement of WZNW models; though, we use
it here in this context for convenience with the intention to present the analysis results of
reference \cite{fh} in a clear way.}
\begin{eqnarray}
h _{(j,m)} = -\frac{j(j+1)}{k-2}+\frac{m^{2}}{k}
\end{eqnarray}
satisfying the Virasoro constraint $h _{(j,m)}=\bar{h}_{(j,\bar{m%
})}=1$ as a requirement of string theory. The quantum numbers
parametrizing the spectrum of strings fall within the lattice in the
following form 
\begin{eqnarray}
p \equiv m+\bar{m}\in \Z \ , \ \ \ \ k\omega \equiv m-\bar{m}\in k\Z  \label{fuld3}
\end{eqnarray}
which can be derived from the level matching condition ({\it cf.} section 2). The quantum numbers $m$ and $\bar m$ are identified with the right momentum $\frac {\sqrt k}{2} P_L$ and the left momentum $\frac {\sqrt k}{2} P_R$ on the cylinder respectively. These can be compared with the case of 2D black hole $SL(2,\R )/U(1)$. 

The central charge of sine-Liouville conformal model is given by 
\begin{equation}
c=2+\frac{6}{k-2}
\end{equation}
Then, the restriction $c=26$ is also required to
define the string theory.

\subsubsection*{The prescription for correlation functions}

As a first step in our analysis, we consider necessary to start by briefly reviewing
the integral construction proposed in \cite{fh}. The treatment of the
sine-Liouville interaction term as a perturbation fits in the philosophy of
the Coulomb gas realization. Actually, the introduction of interaction
effects can be viewed as the insertion of screening charges into the
correlation functions.

Let us define the screening operators by noticing that the interaction term can be rewritten in the following way
\begin{equation}
\frac{\lambda }{4\pi }\int d^{2}ze^{-\sqrt{\frac{k-2}{2}}\phi }\cos \left( 
\sqrt{\frac{k}{2}}\left( X _{L}-X _{R}\right) \right) =\frac{\lambda }{%
8\pi }\int d^{2}z\left( \Phi _{1-\frac{k}{2},\frac{k}{2},-\frac{k}{%
2}} + \Phi _{1-\frac{k}{2},-\frac{k}{2},\frac{k}{2}}  \right)  \label{congo}
\end{equation}
Now, we move to the correlation functions. The Coulomb-gas like prescription
for the correlators applied to the case of sine-Liouville theory leads to
write down the following integral expression for the general $N$-point
functions: 
\begin{eqnarray*}
&&{\mathcal A} _{j_{1}...j_{N}} =\frac{\lambda
^{s_{+}+s_{-}}}{\Gamma (s_{+}+1)\Gamma (s_{-}+1) Vol\left( SL(2,\C)\right) }\prod_{a=1}^{N}\int
d^{2}z_{a}\prod_{a<b}^{N-1,N}\left| z_{a}-z_{b}\right| ^{-\frac{4j_{a}j_{b}}{%
k-2}}\times  \\
&&\ \times \left( z_{a}-z_{b}\right) ^{\frac{2}{k}m_{a}m_{b}}\left( \bar{z}%
_{a}-\bar{z}_{b}\right) ^{\frac{2}{k}\bar{m}_{a}\bar{m}_{b}} \prod_{r=1}^{s_{+}}\int d^{2}u_{r}\prod_{l=1}^{s_{-}}\int
d^{2}v_{l} \times \\
&&\ \times \prod_{a=1}^{N}\left( \prod_{r=1}^{s_{+}}\left| z_{a}-u_{r}\right|
^{2(j_{a}+m_{a})}\left( \bar{z}_{a}-\bar{u}_{r}\right)
^{-p_{a}}\prod_{l=1}^{s_{-}}\left| z_{a}-v_{l}\right|
^{2(j_{a}-m_{a})}\left( \bar{z}_{a}-\bar{v}_{l}\right) ^{p_{a}}\right)
\times   \nonumber  \\
&&\ \times \left( \prod_{r<t}^{s_{+}-1,s_{+}}\left| u_{r}-u_{t}\right|
^{2}\prod_{l<t}^{s_{-}-1,s_{-}}\left| v_{t}-v_{s}\right|
^{2}\prod_{l=1}^{s_{-}}\prod_{r=1}^{s_{+}}\left| v_{l}-u_{r}\right|
^{2-2k}\right)
\end{eqnarray*}
where $d^{2}u_{r}=\frac{du_{r}d\bar{u}_{r}}{2\pi i}$ (resp. $z_{a}$, $w_{l}$%
). This is the standard representation for the perturbative analysis of
conformal models. A factor o the form $\sim \Gamma (-s_+)\Gamma (-s_-)$ could emerge in the functional integration for certain particular {\it naive} prescription for the evaluation of the product of delta functions in the integration over the zero-mode of the $\phi$ field; however, it is necessary to observe that this pair of gamma functions do not indicate the correct pole structure yielding from the integration. These gamma functions before the expression for
residue would be divergent for $s_{\pm }\in \Z_{\geq 0}$. The integration over the zero-modes encodes certain subtelties when two different screening operators are considered and, here, it receives importance in the context of our further discussion about the pole structure; we will consider the corresponding divergent pole contribution yielding in such integration opportunely below.

The charge symmetry conditions, yielding from the integration over the
zero-modes, take the form 
\begin{eqnarray}
\sum_{a=1}^{N}j_{a}+1 &=&\frac{k-2}{2}\left( s_{+}+s_{-}\right)  \label{termo} \ , \ \sum_{a=1}^{N}m_{a} \mp \bar{m}_{a} =\frac{k}{2}\left( s_{+}-s_{-} \pm  s_{+} \mp s_{-}  \right) .  \label{termo2}
\end{eqnarray}
On the other hand, let us mention that the existence of non-trivial zero-dimension fields in the world-sheet theory, namely
\begin{equation}
\Phi _{-\frac{k}{2},\pm \frac{k}{2},\mp \frac{k}{2}%
}(z,\bar{z})= : e^{-\frac{k}{\sqrt{2(k-2)}}\phi (z,\bar{z})\pm i\sqrt{\frac{k%
}{2}}\left( X _{L}(z)-X _{R}(\bar{z})\right) } : \ ,
\end{equation}
could suggest the idea to include such fields in correlators with the purpose to break the chiral balance of $\omega $inding number without introducing a different amount of both screening charges; this idea was considered in references \cite{fzz} and \cite{mo3} within the context of WZNW model. In fact, it is possible to show that if this alternative construction is considered in sine-Liouville model, then differences appear with respect of realization \cite{fh} when the factorization limit is performed.

Here, our intention is to consider the structure of the correlation functions
representing processes violating the total $\omega $inding number for the case of
sine-Liouville string theory as it was originally effectuated in
reference \cite{fh}. Computations of analogous quantities were presented for the case of the $SL(2,\R )_k$ WZNW model in references \cite{nos3,fzz,mo3}. Despite the similarities
between both CFT's, substantial differences between them can be emphasized: an example of these is the role played by the screening operators of the sine-Liouville
CFT and the WZNW model; {\it i.e.} in the case of sine-Liouville theory the violation
of the $\omega $inding number is produced by the presence of the sine-Liouville
interaction term (which is manifestly represented in the free field
realization by the insertion of a different amounts of each screening
operators $s_{+}$ and $s_{-}$), meanwhile, the understanding of the violation of the total $\omega $inding number
in the case of the WZNW model involves a more subtle approach because of the
non-explicit symmetry breaking at the level of the classical action. Moreover, in terms of the string theory on the cigar and string theory on $AdS_3$ the violation ($\delta \omega \neq 0$) is related to the topological properties of these simply connected manifolds.

The comparison between the pole structure of the expression for $\omega $inding violating processes in sine-Liouville theory \cite{fh} and the similar
expression for the non-compact WZNW model becomes a subject of great importance within the
context of the study of FZZ conjecture. The agreement between both computations found here could be other of the
convincing confirmations of the validity of the hypotesis of this strong-weak
duality. We focus the attention on this in the following paragraphs.

Now, let us briefly analyse the computation presented by Fukuda and Hosomichi
in the cited paper. There, the authors have been able to write down a
generic three-point function on the sphere by explicitly evaluating a
modified Dotsenko-Fateev type integral formula. Among other results, they
proved that the $\omega $inding number conservation can be violated up to $\pm 1$
(i.e. $\left| \delta \omega \right| \leq 1$). This is in agreement with
previous calculations effectuated by Zamolodchikov {\it et al.}.

In a very careful analysis of the integral representation, the authors of
reference \cite{fh} have shown that it is feasible to translate the
integrals $\prod_{r,l}\int d^{2}u_{r}\int d^{2}v_{l}$ over the whole complex
plane into the product of contour integrals. Thus, the integral
representation turns out to be described by the techniques
developed in the context of the known integral realization of this class of conformal
field theories. In \cite{fh} these techniques were used and extended in
order to define a precise prescription to explicitly evaluate the
correlators by giving the formula for the contour integrals in the
particular case of sine-Liouville theory. The first step in the calculation
was to decompose the $u_{r}$ complex variables (resp. $v_{l}$) into two
independent real parameters ($\R e\left( u_{r}\right) ,{\baseRing{I} m}\left(
u_{r}\right) $) which take values in the whole real line. Secondly, a Wick
rotation for $\baseRing{I} m(u_{r})$ was peformed in order to introduce a
shifting parameter $\varepsilon $ which is then used to elude the poles in $%
z_{a}$. Then, the contours are taken in such a way that the inserting points $z_{a}$
are avoided by considering the alternative order with respect to these
inserting points. A detailed description for the particular prescription
could be found in the original reference, where the authors refer to the quoted
works by Dotsenko {\it et al.} \cite{dot1,dot3,dot4,dot2}.

The main results of \cite{fh} were the explicit formulas for three-point correlation functions in sine-Liouville CFT, including the cases of $\omega $inding violating processes as well.

Also in \cite{fh}, the obtained expression for correlators in sine-Liouville
theory was compared with the known formula for the three-point function in
the WZNW model on the coset $SL(2,\R )/U(1)$ for the particular case where the
total $\omega $inding number is conserved. The authors showed that the two
expressions are in good agreement in that case. Here, we are interesting in extending the comparison to the $\omega $inding violating correlators.

On the other hand, the comparison of the functional form of 2-point functions in both models
can be studied with detail in the second section of reference \cite{kkk}.
That analysis follows the steps developed in the original work by
Zamolodchikov {\it et al.} There, it is shown how the integration over the
zero-modes of sine-Liouville fields leads to obtain the same analytic
structure of the 2-point function in WZNW model on the cigar manifold.

\subsubsection*{The three-point function violating the winding number}

Again, with the intention to extend this comparison, let us constrast here the
expressions for three-point functions for both theories in the case where
the $\omega $inding number is violated, being $\left| \delta \omega \right| =1$.

First of all, for our purpose, we consider the formula for the evaluation of the residues of the (non-perturbative) three-point functions in sine-Liouville theory \cite{fh}. This is given in terms of $\Upsilon _b $ functions as follows 
\begin{eqnarray}
{\mathcal A} ^{j_1,j_2,j_3}_{m_1,m_2,m_3} &=& \frac {b^{(2b^2 ( 1+\sum_{a=1}^{3}j_{a})-1)(b^{-2}+2)}(-1)^{b^2( 1+\sum_{a=1}^{3}j_{a}+\frac{1}{2b^2} )+m_3 +\bar {m}_3} \pi ^{2b^2 ( 1+\sum_{a=1}^{3}j_{a})}  \Upsilon _b (b^{-1})}{\Gamma ^2\left( b^2\left( 1+\sum_{a=1}^{3}j_{a}+\frac{1}{2b^2} \right) \right) \Upsilon _b \left( b\left( 1+\sum_{a=1}^{3}j_{a}+\frac{1}{2b^2} \right) \right)} \times \nonumber \\
&& \times \prod _{a=1}^{3} \frac {\Gamma (j_a+m_a+1) \Upsilon _b \left( b\left( 2j_{a}+1\right) \right) }{\Gamma (\bar {m}_a -j_a) \Upsilon _b \left( b\left( 2j_a -\sum_{c=1}^{3}j_{c} +\frac{1}{2b^2}\right) \right) }  \nonumber \\ \label{fukuf}
\end{eqnarray}
where $b^{-2}=(k-2)$ and the $\Upsilon _b $ functions are defined by 
\[
\log \Upsilon _b (x)=\frac{1}{4}\int_{0}^{\infty }\frac{d\tau }{\tau }\left(
b+b^{-1}-2x\right) ^{2}e^{-\tau }-\int_{0}^{\infty }\frac{d\tau }{\tau }%
\frac{\sinh ^{2}\left( \frac{\tau }{4}(b+b^{-1}-2x)\right) }{\sinh \left( 
\frac{b\tau }{2}\right) \sinh \left( \frac{b^{-1}\tau }{2}\right) } 
\]
These functions have zeros in the lattice 
\begin{eqnarray}
x \in -b\Z_{\geq 0}-b^{-1}\Z_{\geq 0} \ , \ \ x \in b\Z_{>0}+b^{-1}\Z_{>0}  \label{lcdtm}
\end{eqnarray}
And these\footnote{For a detailed discussion about the analytic properties of $\Upsilon _b $ functions in a related context see \cite{strominger}.} can be written in terms of the special functions $G_{k}$ taking
into account the equivalence 
\begin{equation}
\Upsilon _b ^{-1}(-bx)=b^{b^{2}x^{2}+(1+b^{2})x}G_{b^{-2}+2}(x)  \label{ide}
\end{equation}
Now, let us observe that the formula (\ref{fukuf}) for the residue, obtained originally in \cite{fh}, has to be complemented by including the divergent overall factor $(\sum _{a=1}^{3} j_a +1-\frac{k-2}{2} (s_+ +s_- ))^{-1}$ coming from the integration over the zero-mode of sine-Lioville fields. In fact, the insertion of screening charges in conformal theories with non-compact target space leads to obtain a divergent factor ({\it e.g.} the factors of the form $\sim \Gamma (-s)$ in Liouville CFT); in sine-Liouville theory this factor yields from the integration over the zero-modes of the fields $\phi $ and $X $. Near these poles, the correlation functions take the schematic form\footnote{G.G. is grateful to K. Hosomichi for drawing his attention to this point.}
\[
{\mathcal A} ^{j_1, ...j_N}  = \frac{\lambda ^{s_{+}+s_{-}}\left\langle
\prod_{a=1}^{N}\int d^{2}z_{a}\Phi _{j_{a},m_{a},\bar{m}_{a}}(z_{a},%
\bar{z}_{a})\left( \int d^{2}u\Phi _{1-\frac k2, \frac k2, -\frac k2}(u,\bar{u})\right) ^{s_{+}}\left(
\int d^{2}v\Phi _{1-\frac k2, -\frac k2, \frac k2}(v,\bar{v})\right) ^{s_{-}}\right\rangle _{\lambda
=0}}{\Gamma (s_{+}+1)\Gamma (s_{-}+1)\left( 
\sum_{a=1}^{N}j_{a}+1-\frac{k-2}{%
2}\left( s_{+}+s_{-}\right) \right) }
\]
This implies that the formula should include an additional factor $\sum _{a=1}^{3} j_a +1-\frac{k-2}{2} (s_+ +s_- )$ in the denominator ({\it cf.} \cite{fh}). This overall factor has a crucial role in the pole structure as we will see. For instance, it is easy to observe that, in the particular case $s_-=0$, it leads to the expected overall factor $\Gamma (-s_+)$ and the delta function $\delta \left( \sum_{a=1}^{N}j_{a}+1-\frac{k-2}{2} s_{+} \right)$.

On the other hand, we find convenient to remark that (\ref{ide}) and the functional properties of $G_k$ functions imply 
\[
\Upsilon _b \left( b (\sum _{a=1}^{3} j_a + \frac k2) \right)= \frac { b^{2b^2 (\sum _{a=1}^{3} j_a + \frac k2)-1} \Gamma \left( 1- b^2(\sum _{a=1}^{3}j_a+\frac k2)\right)}{\Gamma \left( b^2(\sum _{a=1}^{3}j_a+\frac k2)\right)}\Upsilon _b \left( b (1+\sum _{a=1}^{3} j_a + \frac k2) \right)
\]
And we also observe that the expansion $\Gamma (\epsilon-n) \sim \frac {(-1)^{n}}{\epsilon \Gamma (1+n)}+{\mathcal O} (1) + {\mathcal O} (\epsilon )$ leads to 
\[
\frac {(-1)^{b^2(\sum _{a=1}^{3}j_a+1)-\frac 12}}{(\sum _{a=1}^{3} j_a +1-\frac{k-2}{2} (s_+ +s_- )) \Gamma ^2(\frac 12+b^2(\sum _{a=1}^{3}j_a+1))} \sim \frac {b^2\Gamma(1-b^2(\sum _{a=1}^{3}j_a+\frac k2))}{\Gamma(b^2(\sum _{a=1}^{3}j_a+\frac k2))} 
\]
where we have identified in the limit $b^{-2}\epsilon=\sum _{a=1}^{3} j_a +1-\frac{k-2}{2} (s_+ +s_- )$ (see equation (\ref{termo})). The last two relations, considered together, permit us to observe that the whole $m$-independent pole structure of the formula (\ref{fukuf}) is codified in four $\Upsilon _b$ functions standing in the denominator, which now lead to the following formula
\begin{eqnarray}
{\mathcal A}  ^{j_1,j_2,j_3}_{m_1,m_2,m_3} = \frac {b^{2\sum_{a=1}^{3}j_{a}(b^{2}+1)+2}(-1)^{m_3 +\bar {m}_3} \pi ^{2b^2 ( 1+\sum_{a=1}^{3}j_{a})}  \Upsilon _b (b^{-1})}{\Upsilon _b \left( b\left( 2+\sum_{a=1}^{3}j_{a}+\frac{1}{2b^2} \right) \right)} \prod _{a=1}^{3} \frac {\Gamma (j_a+m_a+1) \Upsilon _b \left( b\left( 2j_{a}+1\right) \right) }{\Gamma (\bar {m}_a -j_a) \Upsilon _b \left( b\left( 2j_a -\sum_{c=1}^{3}j_{c} +\frac{1}{2b^2}\right) \right) }  \nonumber \\ \label{fuku}
\end{eqnarray}
Then, the three-point function in sine-Liouville theory in the case $\left| \delta \omega \right| =1$ presents poles given by the zeros of these $\Upsilon _b 
$ functions; those singular points are precisely located at (\ref{b2})-(\ref{b8}). Notice that the Weyl reflection is (including the factor we pointed out) closed among this set of pole conditions ({\it cf.} \cite{fh}).

\subsubsection*{Comparing with the WZNW theory}

Now, in analogous way, we can return to the analysis of the pole structure of the three-point
function for $\omega $inding violating processes in the case of the $SL(2,\R )_k /U(1) \times U(1)$ WZNW model. 
We already studied the three-point function for WZNW model in the previous section; and this is given by (\ref{ias}). Then, the positions of the poles are determinated by the analytic
structure of the $G_{k}$ functions, which present poles at (\ref{b2})-(\ref{b8}). And actually, we find an exact agreement between the pole structures of
both (\ref{fuku}) and (\ref{ias}) since the pole conditions (\ref{b2})-(\ref{b8}%
) match exactly with the singular points (\ref{lcdtm}) evaluated for the expression (\ref{ias}).

The crucial point here was the overall factor $\epsilon ^{-1}$. Its inclusion in the formula for sine-Liouville three-point function is necessary to conclude the agreement between both CFT's since, otherwise, a discrepancy between the pole structure of both WZNW and sine-Liouville CFT's would appear due to an additional pole. This is because if one naively compares the formula for the residue of reference \cite{fh} and the correlators computed in \cite{nos3,mo3}, without the consideration of the divergent factor yielding from the integration over the zero-modes of sine-Liouville fields we remarked here, the pole conditions for sine-Liouville which correspond to (\ref{b8}) should be replaced by
\begin{eqnarray*}
-j_{1}-j_{2}-j_{3}-\frac{k}{2} &\in &\Z_{\geq 0}+\Z_{\geq 0}(k-2) \ , \ 1+j_{1}+j_{2}+j_{3}-\frac{k}{2}\in \Z_{\geq 0}+\Z_{\geq 0}(k-2)
\end{eqnarray*}
which does not coincide with the analogous lattice coming from the formula for WZNW and, on the other hand, breaks the Weyl reflection symmetry of the set of pole conditions. This can be seen directly by the $\Upsilon _b$ appearing in equations (\ref{fuku}) and (\ref{fukuf}).

For instance, the poles at $\sum_{a=1}^{3}j_{a}=\frac{k}{2}-2$ arises in
sine-Liouville theory because of the global factor $\sim \Gamma (-s_+)$; and this factor comes from the integration over
the zero-mode. These poles would be at $j_{1}+j_{2}+j_{3}+1\in \Z_{\geq 0}(\frac{k}{2}-1),$
and these are well understood in terms of the integration over the target
space because of its non-compactness.

Besides, we see how the group theoretical factor (\ref{trucoa}) standing in WZNW conformal theory is also in agreement with its conjectured dual model. Strictly speaking, the transformation $\bar m +m \rightarrow \bar m -m$ needs to be considered when comparing with the theory on $SL(2,\R) /U(1)$. This identification is precisely consistent with the definition of ${\omega }$inding number ${\omega }$ and angular momentum $p$ in both theories. 

Let us observe that also the zeros of both expressions (\ref{ias}) and (\ref{fuku}) agree due to Weyl reflection considered for $j_2$ and $j_3$, and the agreement is exact despite the distinctive argument $-2j_2 -k$ of the $G_k$ function in (\ref{ias}). Hence, we conclude that both CFT's are in good agreement when the comparison embraces the $\omega $inding violating three-point correlation functions as well.

The goal of these paragraphs was to address the attention to this agreement since this particular point is an important step in the
checking of the conjectured duality. On the other hand, by inverting the
reasoning, the comparison of both pole structures can be used as a mechanism
for testing the results proposed in the literature as representing
correlators in these theories once the FZZ
conjecture is assumed. Actually, with this original motivation we pointed out here the
main features of the comparison. And, indeed, this comparison allowed us for arguing that the additional pole yielding from the integration over the zero-modes of sine-Liouville fields leads to the correct pole structure in the three-point correlators which violate $\omega $inding number.

\[
\]

G.G. is very grateful to J.M. Maldacena and K. Hosomichi for very useful discussions. He also thanks Simeon Hellerman for conversations and for his interest on this work. D.L. thanks C. N\'u\~nez for many conversations. G.G. was supported by the Institute for Advanced Study IAS and Fundaci\'{o}n Antorchas. D.L. was supported by the Ministerio de Educaci\'on y Ciencias (Spain) through FPU grants.

\end{document}